\documentstyle[12pt,epsf]{article}

\newcommand{\beq}{\begin{equation}}
\newcommand{\eeq}{\end{equation}}
\newcommand{\beqa}{\begin{eqnarray}}
\newcommand{\eeqa}{\end{eqnarray}}
\newcommand{\beqan}{\begin{eqnarray*}}
\newcommand{\eeqan}{\end{eqnarray*}}

\newcommand{\ra}{\rightarrow}

\newcommand{\ben}{\begin{enumerate}}
\newcommand{\een}{\end{enumerate}}
\newcommand{\bfl}{\begin{flushleft}}
\newcommand{\efl}{\end{flushleft}}
\newcommand{\ba}{\begin{array}}
\newcommand{\ea}{\end{array}}
\newcommand{\btab}{\begin{tabular}}
\newcommand{\etab}{\end{tabular}}
\newcommand{\bit}{\begin{itemize}}
\newcommand{\eit}{\end{itemize}}

\newcommand{\cA}{{\cal A}}

\newcommand{\vs}{\vspace}
\newcommand{\hs}{\hspace}

\newcommand{\prepr}[1] {\begin{flushright}  {\bf #1} \end{flushright} \vskip
1.cm}
\newcommand{\titul}[1] {\begin{center}{\Large {\bf #1 } } \end{center}
\vskip 0.8cm}

\newcommand{\autor}[1] {\begin{center}  {\bf \lineskip .3cm #1  }
                        \end{center} }

\newcommand{\lugar}[1] {\begin{center}  {\normalsize \bf \it #1   } 
\end{center}}
\newcounter{muni}

\begin{document}
\hbadness=10000
\pagenumbering{arabic}
\begin{titlepage}
\prepr{
Preprint APCTP/98-22  \\ 
\hs{40mm} KEK-TH-594 \\
\hs{20mm} hep-ph/9810369}
\titul{\bf
Decay Constants $f_{D_s^*}$ and $f_{D_s}$\\
from ${\bar{B}}^0\rightarrow D^+ l^- {\bar{\nu}}$ and
${\bar{B}}^0\rightarrow D^+ D_s^{-(*)}$ Decays}
\autor{ Yong-Yeon Keum
\footnote{Monbushou Research Fellow

Email: keum@apctp.kaist.ac.kr, keum@theory.kek.jp 
} } 
\lugar{ Asia Pacific Center for Theoretical Physics \\
207-43 Cheongrangri Dong, Dongdaemun-Gu, 
Seoul 130-012, Korea }
\lugar{ Theory Group, KEK, Tsukuba, Ibaraki 305-0801 Japan }

\thispagestyle{empty}
\vs{10mm}
\begin{abstract}
\noindent{
We calculate the decay constant of  $D_s$ and $D_s^{*}$ with
$\bar{B}^0 \ra D^{+}\ell^{-}\nu$ and $\bar{B}^0 \ra D^{+}D_s^{-(*)}$ decays.
In our analysis we assume the factorization ansatz and use two
different form factor behaviours (constant and monopole-type)
for $F_0(q^2)$.
We also consider the QCD-penguin contributions in hadronic decays within the NDR
renormalization scheme in a NLO calculation.

We estimate the decay constant of the $D_s$ meson to be $219\pm46$ $MeV$
for (pole/pole)-type form factor and $239\pm50$ $MeV$
for (pole/constant)-type form factor. For $D_s^{*}$ meson, we predict
$f_{D_s^{*}} = 336 \pm 79$ $MeV$, and $f_{D_s^{*}}/f_{D_s} = 1.41 \pm 0.41$
for (pole/constant)-type form factor.
} 

\vs{10mm}
{\rm  PACS index : 12.15.-y, 13.20.-v, 13.25.Hw, 14.40.Nd, 14.65.Fy}

Keywards : Factorization, Non-leptonic Decays, Decay Constant, Penguin Effects
\end{abstract}

\thispagestyle{empty}
\end{titlepage}
\pagebreak

\baselineskip 22pt

\noindent
{\bf \large 1. Introduction}

Measuring purely leptonic decays of heavy mesons provides the cleanest way 
to determine the weak decay constants of heavy mesons, which connect 
the measured quantities, such as the $B\bar{B}$ mixing ratio, to CKM matrix
elements $V_{cb}, V_{ub}$.
However, at present it is not possible to determine $f_B, f_{B_s}$ 
$f_{D_s}$
and
$f_{D^{*}_s}$
experimentally from leptonic $B$ and $D_s$ decays.
For instance, the decay rate for $D_s^{+}$ is given by \cite{rosner}
\beq
\Gamma(D_s^{+} \ra \ell^{+} \nu) = {G_F^2 \over 8 \pi} f_{D_s}^2 m_{\ell}^2 
M_{D_s}
\left( 1 - {m_{l}^2 \over M_{D_s}^2} \right)^2 |V_{cs}|^2
\eeq
Because of helicity suppression, the electron mode $D_s^{+} \ra e^{+}\nu$
has a very small rate. The relative widths are $10 : 1 : 2 \times 10^{-5}$ for
$\tau^{+}\nu, \mu^{+}\nu$ and $e^{+}\nu$ final states, respectively.
Unfortunately the mode with the largest branching fraction, $\tau^{+}\nu$,
has at least two neutrinos in the final state and is difficult to detect experimentally. 
So, theoretical calculations for decay constant have to be used.
The factorization ansatz for nonleptonic decay modes provides us a good
approximate method to obtain nonperturbative quantities, i.e., form factors
and decay constants which are hardly accessible in any other way
\cite{bort-stone,HF}.

There are many ways that the quarks produced in a nonleptonic weak decay can 
arrange
themselves into hadrons. 
The final state is linked to the initial state by complicated trees of gluon and 
quark interactions, pair production, and loops. These make the theoretical 
description of nonleptonic decays difficult.
However, since the products of a $B$ meson decay are quite energetic,
it is possible that the complicated QCD interactions are less important and 
that the two quark pairs of the currents in the weak Hamiltonian,
group individually into the final state mesons without further exchanges of 
gluons.
The color-transparency argument suggests that a quark-antiquark pair remains 
a state of small size with a correspondingly small chromomagnetic moment 
until it is far from the other decay products.

Color transparency is the basis for the factorization hypothesis, in which
amplitudes factorize into products of two current matrix elements.
This ansatz is widely used in heavy quark physics, 
as it is almost the only way to treat hadronic decays.

In this paper we consider a way to determine weak decay constants
$f_{D_s}$ and $f_{D_s^{*}}$ under factorization ansatz.
In section 2, we discuss how to extract the unknown parameter
$|V_{cb}F_1^{BD}(0)|$ from the branching ratio of the semileptonic decay
$\bar{B}^{0} \ra D^{+}\ell \bar{\nu}$.
In order to check the validity of the factorization assumption,
we study the nonleptonic two body decays, for instance,
$B \ra D\rho, D\pi$ and $DK^{(*)}$ in section 3.
In section 4, we discuss the determination of $f_{D_s}$
and $f_{D_s^{*}}$ from $\bar{B}^0 \ra D^{+}D_s^{-(*)}$ decay modes.
In our analysis, we consider the QCD-penguin effects 
 which amount to
about 11 $\%$ for $B \ra D D_s$ and 5 $\%$ for $B \ra D D_s^{*}$,
which have been mentioned in Ref.\cite{GKYP}. 

\vs{15mm}
{\bf \large 2. Semileptonic Decay
${\bar{B}}^0\rightarrow D^+ l^- {\bar{\nu}}$} 

From Lorentz invariance one finds the decomposition of the
hadronic matrix element in terms of hadronic form factors:
\begin{eqnarray}
<D^+(p_D )|J_\mu |{\bar{B}}^0(p_B)>
&=&
\left[ (p_B+p_D )_\mu 
-{m_B^2-m_D^2\over q^2}q_\mu \right] \, F_1^{BD}(q^2)
\nonumber \\
\cr
&&
\hs{3mm}
+{m_B^2-m_D^2\over q^2}\, q_\mu \, F_0^{BD}(q^2),
\label{a1}
\end{eqnarray}
where $J_\mu = {\bar{c}}\gamma_\mu b$ and
$q_\mu =(p_B-p_D )_\mu$.
In the rest frame of the decay products, $F_1(q^2)$ and $F_0(q^2)$
correspond to $1^-$ and $0^+$ exchanges, respectively.
At $q^2=0$ we have the constraint
\begin{equation}
F_1^{BD}(0)=F_0^{BD}(0),
\label{a1a}
\end{equation}
since the hadronic matrix element in (\ref{a1}) is nonsingular
at this kinematic point.

The $q^2$ distribution in the semileptonic decay
${\bar{B}}^0\rightarrow D^+ l^- {\bar{\nu}}$
is written in terms of the hadronic form factor
$F_1^{BD}(q^2)$ as
\begin{equation}
{d\Gamma ({\bar{B}}^0\rightarrow D^+ l^- {\bar{\nu}})\over dq^2}
={G_F^2\over 24\pi^3}\, |V_{cb}|^2\, [K(q^2)]^3\,
|F_1(q^2)|^2,
\label{a2}
\end{equation}
where the $q^2$ dependent momentum $K(q^2)$ is given by
\begin{equation}
K(q^2)={1\over 2m_B}\,
\left[ (m_B^2+m_D^2-q^2)^2
-4m_B^2m_D^2 \right]^{1/2}.
\label{a3}
\end{equation}
In the zero lepton-mass limit,
$0\le q^2\le (m_B-m_D)^2$.

For the $q^2$ dependence of the form factors,
Wirbel $et$ $al.$ \cite{wsb} assumed a simple pole formula
for both $F_1(q^2)$ and $F_0(q^2)$ (we designate this scenario 'pole/pole'):
\begin{equation}
F_1(q^2)=F_1(0)\,/ ( 1-{q^2\over m_{F_1}^2}),\qquad
F_0(q^2)=F_0(0)\,/ ( 1-{q^2\over m_{F_0}^2}),
\label{a4}
\end{equation}
with the pole masses
\begin{equation}
m_{F_1}=6.34\ {\rm GeV},\qquad
m_{F_0}=6.80\ {\rm GeV}.
\label{a4aa}
\end{equation}
Korner and Schuler \cite{ks} have also adopted the same $q^2$ dependence of
$F_1(q^2)$ and $F_0(q^2)$ given by (\ref{a4}) and (\ref{a4aa}).
On the other hand, the heavy quark effective theory (HQET)
gives in the $m_{b,c}\rightarrow \infty$ limit
the relation between $F_1(q^2)$ and $F_0(q^2)$
given by \cite{iw,nr}
\begin{equation}
F_0(q^2)=\left[ 1-{q^2\over (m_B+m_D)^2}\right] \hs{3mm} F_1(q^2).
\label{a4a1}
\end{equation}
The combination of (\ref{a4}) and (\ref{a4a1}) suggests that $F_0(q^2)$
is approximately constant when we keep the simple pole dependence for
$F_1(q^2)$.
Therefore, in this paper, as well as the above 'pole/pole' form factors,
we will also consider the following ones (designated 'pole/const.'):
\begin{equation}
F_1(q^2)=F_1(0)\,/( 1-{q^2\over m_{F_1}^2}),\qquad
F_0(q^2)=F_0(0),
\label{a4b1}
\end{equation}
with
\begin{equation}
m_{F_1}=6.34\ {\rm GeV}.
\label{a4aab2}
\end{equation}

By introducing the variable
$x\equiv q^2/m_B^2$,
which has the range of
$0\le x\le (1-{m_D\over m_B})^2$
in the zero lepton mass limit,
(\ref{a2}) is written as
\begin{eqnarray}
{d\Gamma ({\bar{B}}^0\rightarrow D^+ l^- {\bar{\nu}})\over dx}
&=&{G_F^2m_B^5\over 192\pi^3}\, |V_{cb}\, F_1^{BD}(0)|^2\,
{{\lambda}^{3}[1,\, {m_D^2\over m_B^2},\, x] \over
\Bigl( 1-{m_B^2\over m_{F_1}^2}x{\Bigr)}^2},
\label{a5}\\
{\lambda}[1,\, {m_D^2\over m_B^2},\, x]&=&
\left[ (1+{m_D^2\over m_B^2}-x)^2-4{m_D^2\over m_B^2}
\right]^{1/2}.
\nonumber
\end{eqnarray}
Then the branching ratio
${\cal{B}} ({\bar{B}}^0\rightarrow D^+ l^- {\bar{\nu}})$
is given by
\begin{eqnarray}
{\cal{B}} ({\bar{B}}^0\rightarrow D^+ l^- {\bar{\nu}})&=&
({G_Fm_B^2\over {\sqrt{2}}})^2\, {m_B\over {\Gamma}_B}\,
{2\over 192 {\pi}^2}\,
|V_{cb}\, F_1^{BD}(0)|^2\times I
\nonumber\\
&=&
2.221\times 10^2 \,\, |V_{cb}\, F_1^{BD}(0)|^2\times I,
\label{a6}
\end{eqnarray}
where the dimensionless integral $I$ is given by
\begin{equation}
I=
\int_0^{(1-{m_D\over m_B})^2} dx
{ \left[ (1+{m_D^2\over m_B^2}-x)^2
-4{m_D^2\over m_B^2} \right]^{3/2}\over
\Bigl( 1-{m_B^2\over m_{F_1}^2}x{\Bigr)}^2}
=0.121
\label{a7}
\end{equation}
In obtaining the numerical values in (\ref{a6}) and (\ref{a7}),
we have used the following
experimetal results \cite{rpp}:
$m_D=m_{D^+}=1.869$ GeV, $m_B=m_{B^0}=5.279$ GeV,
${\Gamma}_B={\Gamma_{B^0}}=4.219\times 10^{-13}$ GeV
(${\tau}_{B^0}=(1.56\pm 0.06)\times 10^{-12}$ s),
and $G_F=1.166\ 39(2)\times 10^{-5}\ {\rm GeV}^{-2}$.
Since
${\cal{B}} ({\bar{B}}^0\rightarrow D^+ l^- {\bar{\nu}})=
(1.78\pm 0.20\pm 0.24)\times 10^{-2}$ is obtained experimentally
\cite{stone},
the value of $|V_{cb}\, F_1^{BD}(0)|$ can be extracted from (\ref{a6}).
Following this procedure, we obtain 
\begin{equation}
|V_{cb}\, F_1^{BD}(0)|=(2.57\pm 0.14\pm 0.17)\times 10^{-2}.
\label{a7a}
\end{equation}
In the calculations that follow,
we will use $|V_{cb}\, F_1^{BD}(0)|=(2.57\pm 0.22)\times 10^{-2}$
which combines the statistical and systematic errors in
(\ref{a7a}).

\vs{15mm}
{\bf \large 3. Test of Factorization with
${\bar{B}}^0\rightarrow D^+ \rho^-$ and
${\bar{B}}^0\rightarrow D^+ \pi^-$,
and Prediction of Branching Ratio
${\cal{B}}({\bar{B}}^0\rightarrow D^+ K^{-(*)})$
} \\

We start by recalling the relevant effective weak Hamiltonian:
\begin{equation}
{\cal H}_{\rm eff}={G_F\over {\sqrt{2}}}V_{cb}V_{ud}^*
[C_1(\mu ){\cal O}_1+C_2(\mu ){\cal O}_2]\ +\ {\rm H.C.},
\label{b1}
\end{equation}
\begin{equation}
{\cal O}_1=({\bar{d}}\Gamma^\rho u)({\bar{c}}\Gamma_\rho b),\quad
{\cal O}_2=({\bar{c}}\Gamma^\rho u)({\bar{d}}\Gamma_\rho b),
\label{b2}
\end{equation}
where $G_F$ is the Fermi coupling constant, $V_{cb}$ and $V_{ud}$
are corresponding Cabibbo-Kobayashi-Maskawa (CKM) matrix elements
and $\Gamma_\rho = \gamma_\rho (1-\gamma_5)$.
The Wilson coefficients $C_1(\mu )$ and $C_2(\mu )$ incorporate
the short-distance effects arising from the scaling of
${\cal H}_{\rm eff}$ from $\mu =m_W$ to $\mu =O(m_b)$.
By using the Fierz transformation under which $V-A$ currents remain
$V-A$ currents, we get the following equivalent forms:
\begin{eqnarray}
C_1{\cal O}_1+C_2{\cal O}_2&=&
(C_1+{1\over N_c}C_2){\cal O}_1
+C_2({\bar{d}}\Gamma^\rho T^au)({\bar{c}}\Gamma_\rho T^ab)
\nonumber\\
&=&(C_2+{1\over N_c}C_1){\cal O}_2
+C_1({\bar{c}}\Gamma^\rho T^au)({\bar{d}}\Gamma_\rho T^ab),
\label{b3}
\end{eqnarray}
where $N_c=3$ is the number of colors and $T^a$s are $SU(3)$ color
generators.
The second terms in (\ref{b3}) involve color-octet currents.
In the factorization assumption, these terms are neglected and
${\cal H}_{\rm eff}$ is rewritten in terms of ``factorized hadron
operators'' \cite{wsb,keum}:
\begin{equation}
{\cal H}_{\rm eff}={G_F\over {\sqrt{2}}}V_{cb}V_{ud}^*
\Bigl( a_1[{\bar{d}}\Gamma^\rho u]_H[{\bar{c}}\Gamma_\rho b]_H
+a_2[{\bar{c}}\Gamma^\rho u]_H[{\bar{d}}\Gamma_\rho b]_H\Bigr)
\ +\ {\rm H.C.},
\label{b4}
\end{equation}
where the subscript $H$ stands for $hadronic$ implying that the
Dirac bilinears inside the brackets be treated as interpolating
fields for the mesons and no further Fierz-reordering need be done.
The phenomenological parameters $a_1$ and $a_2$ are related to
$C_1$ and $C_2$ by
\begin{equation}
a_1=C_1+{1\over N_c}C_2,\quad
a_2=C_2+{1\over N_c}C_1.
\label{b5}
\end{equation}
From the analyses of A.J. Buras \cite{burasNDR} , the parameters $a_1$ and
$a_2$ are determined in the Next-to-Leading-Order (NLO) calculation in the Naive Dimensional Regularization (NDR) scheme as
\begin{equation}
a_1=1.02\pm 0.01,\quad a_2=0.20\pm 0.05.
\label{b6}
\end{equation}

For the two body decay, in the rest frame of initial meson
the differential decay rate is given by
\begin{equation}
d\Gamma ={1\over 32\pi^2}|{\cal M}|^2
{|{\bf p}_1|\over M^2}d\Omega ,
\label{b7}
\end{equation}
\begin{equation}
|{\bf p}_1|=
{[(M^2-(m_1+m_2)^2)(M^2-(m_1-m_2)^2)]^{{1/2}}\over 2M},
\label{b8}
\end{equation}
where $M$ is the mass of initial meson, and $m_1$ ($m_2$) and
${\bf p}_1$ are the mass and momentum of one of final mesons.
By using (\ref{a1}), (\ref{b4}) and
$<0|\Gamma_\mu |\rho(q,\varepsilon )>
=\varepsilon_\mu (q) m_{\rho} f_{\rho}$,
(\ref{b7}) gives the following formula for the branching ratio of
${\bar{B}}^0\rightarrow D^+ \rho^-$:
\begin{eqnarray}
{\cal{B}} ({\bar{B}}^0\rightarrow D^+ \rho^-)
&=& \left({G_Fm_B^2\over {\sqrt{2}}}\right)^2\,
|V_{ud}|^2\, {1\over 16 \pi }\, {m_B\over {\Gamma}_B}\,
a_1^2\,
{f_{\rho}^2 \over m_B^2}\, |V_{cb}\, F_1^{BD}(m_{\rho}^2)|^2
\nonumber\\
&\times &
\left[\Bigl( 1-({m_D+m_{\rho}\over m_B})^2{\Bigr)}
 \Bigl( 1-({m_D-m_{\rho}\over m_B})^2{\Bigr)}
\right]^{3/2}
\nonumber\\
&=&13.25
\times |V_{cb}\, F_1^{BD}(m_{\rho}^2)|^2.
\label{b12}
\end{eqnarray}
In obtaining the numerical values in (\ref{b12}), we have used $m_{\rho}=m_{\rho^+}=766.9$ MeV,
$f_{\rho}=f_{\rho^+}=216$ MeV,
$V_{ud}=0.9751$ \cite{rpp}, and the
parameters given below (\ref{a7}).

For  $a_1$ we use the value given in (\ref{b6}).
Then, by using the formula (\ref{b12}) with the values
of $|V_{cb}\, F_0^{BD}(0)|^2$, $(F_0^{BD}(0)=F_1^{BD}(0))$
given in (\ref{a7a}),
we obtain the branching ratio
${\cal{B}}({\bar{B}}^0\rightarrow D^+ \rho^-)$ presented in Table 3.

For the process ${\bar{B}}^0\rightarrow D^+ K^{*-}$,
by using
$<0|\Gamma_\mu |K^{*}(q,\varepsilon )>
=\varepsilon_\mu (q) m_{K^{*}} f_{K^{*}}$,
we have
\begin{eqnarray}
{\cal{B}} ({\bar{B}}^0\rightarrow D^+ K^{*-})
&=& \left({G_Fm_B^2\over {\sqrt{2}}}\right)^2\,
|V_{us}|^2\, {1\over 16\pi }\, {m_B\over {\Gamma}_B}\,
a_1^2\,
{f_{K^*}^2 \over m_B^2}\, |V_{cb}\, F_1^{BD}(m_{K^*}^2)|^2
\nonumber\\
&\times &
\left[\Bigl( 1-({m_D+m_{K^*}\over m_B})^2{\Bigr)}
 \Bigl( 1-({m_D-m_{K^*}\over m_B})^2{\Bigr)}
\right]^{3/2}
\nonumber\\
&=&0.67\times
|V_{cb}\, F_1^{BD}(m_{K^*}^2)|^2.
\label{b12a}
\end{eqnarray}
where we have used $m_{K^*}=m_{K^{*-}}=891.59$ MeV,
$f_{K^*}=f_{K^{*-}}=218$ MeV,
and $V_{us}=0.2215$ \cite{rpp}.
By using (\ref{b12}) with $|V_{cb}\, F_1^{BD}(0)|^2$ in (\ref{a7a}),
we obtain the branching ratio
${\cal{B}}({\bar{B}}^0\rightarrow D^+ K^{*-})$ presented in Table \ref{table3}.

By using (\ref{a1}), (\ref{b4}) and
$<0|\Gamma_\mu |\pi(q)>=iq_\mu f_{\pi}$,
(\ref{b7}) gives the following formula for the branching ratio for the
process ${\bar{B}}^0\rightarrow D^+ \pi^-$:
\begin{eqnarray}
{\cal{B}} ({\bar{B}}^0\rightarrow D^+ \pi^-)
&=&\left({G_Fm_B^2\over {\sqrt{2}}}\right)^2\,
|V_{ud}|^2\, {1\over 16 \pi }\, {m_B\over {\Gamma}_B}\, a_1^2\,
{f_{\pi}^2 \over m_B^2}\, |V_{cb}\, F_0^{BD}(m_{\pi}^2)|^2
\nonumber\\
& &\times \Bigl( 1-{m_D^2\over m_B^2}{\Bigr)}^2\,
\left[\Bigl( 1-({m_D+m_{\pi}\over m_B})^2{\Bigr)}
\Bigl( 1-({m_D-m_{\pi}\over m_B})^2{\Bigr)}\right]^{1/2}
\nonumber\\
&=&5.42\times
|V_{cb}\, F_0^{BD}(m_{\pi}^2)|^2.
\label{b11}
\end{eqnarray}
where we have used
$m_{\pi}=m_{\pi^-}=139.57$ MeV and
$f_{\pi}=f_{\pi^-}=131.74$ MeV \cite{rpp}.
By using the formula (\ref{b11}) with the values
of $|V_{cb}\, F_0^{BD}(0)|^2$, $(F_0^{BD}(0)=F_1^{BD}(0))$
in (\ref{a7a}),
we obtain the branching ratio for 
${\bar{B}}^0\rightarrow D^+ \pi^-$ presented in Table \ref{table3}.

For the process ${\bar{B}}^0\rightarrow D^+ K^-$,
by using $<0|\Gamma_\mu |K^-(q)>=iq_\mu f_{K^-}$,
we have
\begin{eqnarray}
{\cal{B}} ({\bar{B}}^0\rightarrow D^+ K^-)
&=& \left({G_Fm_B^2\over {\sqrt{2}}}\right)^2\,
|V_{us}|^2\, {1\over 16\pi }\, {m_B\over {\Gamma}_B}\, a_1^2\,
{f_{K}^2 \over m_B^2}\, |V_{cb}\, F_0^{BD}(m_{K}^2)|^2
\nonumber\\
& &\times \Bigl( 1-{m_D^2\over m_B^2}{\Bigr)}^2\,
\left[\Bigl( 1-({m_D+m_{K}\over m_B})^2{\Bigr)}
\Bigl( 1-({m_D-m_{K}\over m_B})^2{\Bigr)}
\right]^{1/2}
\nonumber\\
&=&0.41\times
|V_{cb}\, F_0^{BD}(m_{K}^2)|^2,
\label{b11a}
\end{eqnarray}
where we have used $m_{K}=m_{K^-}=493.68$ MeV,
$f_{K}=f_{K^+}=160.6 MeV$ \cite{rpp}.
By using (\ref{b12}) with $|V_{cb}\, F_1^{BD}(0)|^2$ in (\ref{a7a}),
we obtain the branching ratio
${\cal{B}}({\bar{B}}^0\rightarrow D^+ K^-)$ presented in Table \ref{table3}.
An inspection of Table 1 shows that the factorization method works well in $\bar{B}^0 \rightarrow
D^{+}\pi^{-}, D^{+}\rho^{-}$ decays. We predict branching ratios :
\begin{eqnarray} 
&& {\cal B}(\bar{B}^0 \rightarrow D^{+}K^{-}) \simeq 2.7 \cdot 10^{-4}
\nonumber \\
&& {\cal B}(\bar{B}^0 \rightarrow D^{+}K^{*-}) \simeq 4.6 \cdot 10^{-4}
\end{eqnarray}
which is certainly reachable in near future.

\vs{15mm}
{\bf \large 4. Determination of $f_{D_s^{*}}$
and $f_{D_s}$ from
${\bar{B}}^0\rightarrow D^+ D_s^{-*}$ and
${\bar{B}}^0\rightarrow D^+ D_s^-$} \\

{\bf (a) At Tree level}

In this section we calculate the vector and pseudoscalar decay
constants $f_{D_s^*}$ and $f_{D_s}$ from
the experimentally obtained
branching ratios of the the exclusive decays
${\bar{B}}^0\rightarrow D^+ D_s^{-*}$ and
${\bar{B}}^0\rightarrow D^+ D_s^-$.
By using (\ref{a1}), (\ref{b4}) and
$<0|\Gamma_\mu |D_s^{ *}(q,\varepsilon )>
=\varepsilon_\mu (q) m_{D_s^{ *}} f_{D_s^{ *}}$,
(\ref{b7}) gives the following formula for the branching ratio of
${\bar{B}}^0\rightarrow D^+ D_s^{-*}$:
\begin{eqnarray}
{\cal{B}} ({\bar{B}}^0\rightarrow D^+ D_s^{-*})
&=& \left({G_Fm_B^2\over {\sqrt{2}}}\right)^2\,
|V_{cb}|^2\, |V_{cs}|^2\, {1\over 16\pi }\, {m_B\over {\Gamma}_B}\,
a_1^2\,
{f_{D_s^{*}}^2 \over m_B^2}\, |F_1^{BD}(m_{D_s^{*}}^2)|^2
\nonumber\\
&\times &
\left[ \Bigl( 1-({m_D+m_{D_s^{*}}\over m_B})^2{\Bigr)}
 \Bigl( 1-({m_D-m_{D_s^{*}}\over m_B})^2{\Bigr)}
\right]^{3/2}
\nonumber\\
&=&(1.32\cdot 10^2\ {\rm GeV}^2)\cdot f_{D_s^{*}}^2\cdot
|V_{cb}\, F_1^{BD}(m_{D_s^{*}}^2)|^2,
\label{cb12}
\end{eqnarray}
where we have used
$m_{D_s^{*}}=m_{D_s^{-*}}=2112.4$ MeV
$V_{cs}=0.9743$ \cite{rpp}.
Then, from (\ref{cb12}) we get
\begin{equation}
f_{D_s^{*}}=(0.87\times 10^{-1}\ {\rm GeV})\cdot
{\sqrt{{\cal{B}}({\bar{B}}^0\rightarrow D^+ D_s^{-*})}
\over
|V_{cb}\, F_1^{BD}(m_{D_s^{*}}^2)|}.
\label{cb12a}
\end{equation}

For the process ${\bar{B}}^0\rightarrow D^+ D_s^-$,
by using $<0|\Gamma_\mu |D_s(q)>=iq_\mu f_{D_s}$,
we have
\begin{eqnarray}
{\cal{B}} ({\bar{B}}^0\rightarrow D^+ D_s^-)
&=&\left({G_Fm_B^2\over {\sqrt{2}}}\right)^2\,
|V_{cb}|^2\, |V_{cs}|^2\, {1\over 16\pi }\, {m_B\over {\Gamma}_B}\, a_1^2\,
{f_{D_s}^2 \over m_B^2}\, |F_0^{BD}(m_{D_s}^2)|^2
\nonumber\\
& &\times \Bigl( 1-{m_D^2\over m_B^2}{\Bigr)}^2\,
\left[ \Bigl( 1-({m_D+m_{D_s}\over m_B})^2{\Bigr)}
\Bigl( 1-({m_D-m_{D_s}\over m_B})^2{\Bigr)}
\right]^{1/2}
\nonumber\\
&=&(2.45\cdot 10^2\ {\rm GeV}^2)\cdot f_{D_s}^2\cdot
|V_{cb}\, F_0^{BD}(m_{D_s}^2)|^2,
\label{cb11}
\end{eqnarray}
where we used $m_{D_s}=m_{D_s^-}=1968.5$ MeV \cite{rpp}.
From (\ref{cb11}) we get
\begin{equation}
f_{D_s}=(0.64\times 10^{-1}\ {\rm GeV})\cdot
{\sqrt{{\cal{B}}({\bar{B}}^0\rightarrow D^+ D_s^{-})}
\over
|V_{cb}\, F_0^{BD}(m_{D_s}^2)|}.
\label{cb11a}
\end{equation}
Browder $et$ $al.$ \cite{browder} present the following experimental
results for the branching ratios:
\begin{eqnarray}
{\cal{B}}({\bar{B}}^0\rightarrow D^+ D_s^{-*})&=&
(1.14\pm 0.42\pm 0.28)\times 10^{-2}
=(1.14\pm 0.50)\times 10^{-2},
\nonumber\\
{\cal{B}}({\bar{B}}^0\rightarrow D^+ D_s^{-})&=&
(0.74\pm 0.22\pm 0.18)\times 10^{-2}
=(0.74\pm 0.28)\times 10^{-2},
\label{c21}
\end{eqnarray}
where we have combined the statistical and systematic errors.
By using (\ref{c21}) and the values of
$|V_{cb}\, F_1^{BD}(m_{D_s^{*}}^2)|$ and
$|V_{cb}\, F_0^{BD}(m_{D_s}^2)|$ given in Table 2,
from (\ref{cb11a}) and (\ref{cb12a}) we obtain the the following results:
\begin{eqnarray}
f_{D_s^*}=322\pm 76\ {\rm MeV},\qquad f_{D_s}=196\pm 41\ {\rm MeV}& &
{\rm for\ (pole/pole)},
\nonumber\\
f_{D_s^*}=322\pm 76\ {\rm MeV},\qquad f_{D_s}=214\pm 45\ {\rm MeV}& &
{\rm for\ (pole/const.)}.
\label{c22}
\end{eqnarray}
The ratio of the vector and pseudoscalar decay constants
$f_{D_s^*}/f_{D_s}$ is given by
\begin{equation}
{f_{D_s^*}\over f_{D_s}}=1.36 \times
{|V_{cb}\, F_0^{BD}(m_{D_s}^2)|\over |V_{cb}\, F_1^{BD}(m_{D_s^{*}}^2)|}
\times
\left[
{{\cal{B}}({\bar{B}}^0\rightarrow D^+ D_s^{-*})\over
{\cal{B}}({\bar{B}}^0\rightarrow D^+ D_s^{-})} \right]^{{1/2}},
\label{c23}
\end{equation}
which gives
\begin{eqnarray}
{f_{D_s^*}\over f_{D_s}}=1.64\pm 0.48& &{\rm for\ (pole/pole)},
\nonumber\\
{f_{D_s^*}\over f_{D_s}}=1.51\pm 0.44& &{\rm for\ (pole/const.)}.
\label{c24}
\end{eqnarray}

\vs{10mm}
{\bf (b) Including Penguin contribution}

The effective Hamiltonian for $\triangle B = 1$ transitions is given by
\begin{equation}
H_{\rm eff}=
{G_F\over {\sqrt{2}}}[V_{ub}V_{uq}^*(C_1O_1^u+C_2O_2^u)
+V_{cb}V_{cq}^*(C_1O_1^c+C_2O_2^c)
-V_{tb}V_{tq}^*\sum_{i=3}^6C_iO_i],
\label{pb1}
\end{equation}
where $q=d,s$ and $C_i$ are the Wilson coefficients evaluated at the
renormalization scale $\mu$, and
the current-current operators $O_1^{u,c}$ and $O_2^{u,c}$ are
\begin{eqnarray}
O_1^u=({\bar{u}}_\alpha b_\alpha )_{V-A}({\bar{q}}_\beta u_\beta )_{V-A}
&\qquad &
O_1^c=({\bar{c}}_\alpha b_\alpha )_{V-A}({\bar{q}}_\beta c_\beta )_{V-A}
\nonumber\\
O_2^u=({\bar{u}}_\beta b_\alpha )_{V-A}({\bar{q}}_\alpha u_\beta )_{V-A}
&\qquad &
O_2^c=({\bar{c}}_\beta b_\alpha )_{V-A}({\bar{q}}_\alpha c_\beta )_{V-A},
\label{pb2}
\end{eqnarray}
and the QCD penguin operators $O_3\ -\ O_6$ are
\begin{eqnarray}
O_3=({\bar{q}}_\alpha b_\alpha )_{V-A}
\sum_{q'}({\bar{q}}'_\beta q'_\beta )_{V-A}
&\qquad &
O_4=({\bar{q}}_\beta b_\alpha )_{V-A}
\sum_{q'}({\bar{q}}'_\alpha q'_\beta )_{V-A}
\nonumber\\
O_5=({\bar{q}}_\alpha b_\alpha )_{V-A}
\sum_{q'}({\bar{q}}'_\beta q'_\beta )_{V+A}
&\qquad &
O_6=({\bar{q}}_\beta b_\alpha )_{V-A}
\sum_{q'}({\bar{q}}'_\alpha q'_\beta )_{V+A}.
\label{pb3}
\end{eqnarray}
In (\ref{pb1}) we have neglected the effects of the electroweak penguin operators 
and the dipole operators, since their contributions are not important for the
calculations of this paper.

When we take $m_t=174$ GeV, $m_b=5.0$ GeV, $\alpha_{\rm s}(M_z)=0.118$
and $\alpha_{\rm em}(M_z)=1/128$, the numerical values of the renormalization
scheme independent Wilson coefficients ${\bar{C}}_i$ at $\mu =m_b$ are
given by \cite{deshpande}
\begin{eqnarray}
&&{\bar{C}}_1=-0.3125,\ \ \ {\bar{C}}_2=1.1502,
\nonumber\\
&&{\bar{C}}_3=0.0174,\ \ \ {\bar{C}}_4=-0.0373,\ \ \
{\bar{C}}_5=0.0104,\ \ \ {\bar{C}}_6=-0.0459.
\label{pb4}
\end{eqnarray}
The effective Hamiltonian in (\ref{pb1}) for the decays
${\bar{B}}^0\rightarrow D^+ D_s^{-(*)}$
can be rewritten as
\begin{equation}
H_{\rm eff}=
{G_F\over {\sqrt{2}}}
[V_{cb}V_{cs}^*(C_1^{\rm eff}O_1^c+C_2^{\rm eff}O_2^c)
-V_{tb}V_{ts}^*\sum_{i=3}^6C_i^{\rm eff}O_i],
\label{pb5}
\end{equation}
where $C_i^{\rm eff}$ are given by \cite{fleischer}
\begin{eqnarray}
&&C_1^{\rm eff}={\bar{C}}_1,\ \ \ C_2^{\rm eff}={\bar{C}}_2,\ \ \
C_3^{\rm eff}={\bar{C}}_3-P_{\rm s}/N_{\rm c},\ \ \
C_4^{\rm eff}={\bar{C}}_4+P_{\rm s},
\nonumber\\
&&C_5^{\rm eff}={\bar{C}}_5-P_{\rm s}/N_{\rm c},\ \ \
C_6^{\rm eff}={\bar{C}}_6+P_{\rm s}
\label{pb6}
\end{eqnarray}
with
\begin{eqnarray}
&&P_{\rm s}={\alpha_{\rm s}\over 8\pi }
[{10\over 9}-G(m_q,q^2,\mu )]{\bar{C}}_2(\mu ),
\nonumber\\
&&G(m_q,q^2,\mu )=-4\int_0^1x(1-x)\, 
{\rm ln}({m_q^2-x(1-x)q^2\over \mu^2})dx,
\label{pb7}
\end{eqnarray}
where $q$ denotes the momentum of the virtual gluons appearing in the QCD 
time-like matrix elements, and $N_{\rm c}$ is the number of colors.
Assuming $q^2 = m_b^2/2$, we obtain the analytic formular for $G(m_q,q^2,\mu)$ :
\beq
G(m_q,{m_b^2 \over 2},\mu) = -{2 \over 3} ln\left({y \over 8} \right) + {10 
\over 9}
+{2 \over 3} y + {(2 + y)\sqrt{1 - y} \over 3}
\left[ ln\left|{1 - \sqrt{1 - y} \over 1 + \sqrt{1 - y}} \right| + i \pi \right]
\label{pb8}
\eeq
with $y = 8 m_q^2/m_b^2$. By inserting the values for $m_q = m_c = 1.45$ $GeV$,
and $N_c = 3$ into (\ref{pb6}) and (\ref{pb8}),
we get the values $C_{i}^{eff} (i = 1 \sim 6)$ which is shown in Table 
\ref{table4}.

The decay amplitude
${\cal A}_{\rm T+P}({\bar{B}}^0\rightarrow D^+ D_s^-)
=<D^+ D_s^-|{\cal{H}}_{\rm eff}|{\bar{B}}^0>$
is given as follows:
\beqa
{\cal A}_{\rm T+P}({\bar{B}}^0\rightarrow D^+ D_s^-)
&=& {G_F\over {\sqrt{2}}}
[V_{cb}V_{cs}^*a_2 - V_{tb}V_{ts}^*( a_4
+2a_6 {m_{D_s}^2\over (m_b-m_c)(m_s+m_c)})] \hs{2mm} {\cal M}_{a}
\nonumber \\
\cr
& \simeq &
{G_F\over {\sqrt{2}}} V_{cb}V_{cs}^*a_2
\left[1 + {a_4 \over a_2} + 2 {a_6 \over a_2}
{m_{D_s}^2 \over (m_b-m_c)(m_s+m_c)} \right] \hs{2mm}  {\cal M}_{a} 
\label{pb9}
\eeqa
where 
\beq
{\cal M}_a =  <D_s^-|{\bar{s}}\gamma^\mu\gamma_5c|0>
<D^{+}|{\bar{c}}\gamma_\mu b|{\bar{B}}^0>
= - i f_{D_s}(m_B^2 - m_D^2) F_0^{BD}(m_{D_s}^2)
\eeq 

On the other hand, we have
\beqa
{\cal A}_{\rm T+P}({\bar{B}}^0\rightarrow D^+ D_s^{-*})
&=& {G_F\over {\sqrt{2}}}
[V_{cb}V_{cs}^*a_2 -V_{tb}V_{ts}^*a_4] \hs{2mm}
{\cal M}_b \nonumber \\
\cr
& \simeq  &  {G_F\over {\sqrt{2}}}
 V_{cb}V_{cs}^*a_2
\left(1 + {a_4 \over a_2} \right) \hs{2mm} {\cal M}_b
\label{pb10}
\eeqa
where 
\beq
{\cal M}_b =  <D_s^{*}|{\bar{s}}\gamma^\mu\gamma_5c|0>
<D^{+}|{\bar{c}}\gamma_\mu b|{\bar{B}}^0>
= m_{D_s^{*}}f_{D_s^{*}}[\epsilon(q) \cdot (P_B + P_D)] F_1^{BD}(m_{D_s^{*}}^2)
\eeq 
We can estimate the penguin contributions for each process from (\ref{pb9})
and (\ref{pb10}) as follows :
\beqa
{\rm For} \hs{3mm} \bar{B}^0 \ra D^{+} D_s^{-} ; \hs{10mm} & &
\left| {\cA_P \over \cA_T} \right| =
\left|{a_4 \over a_2} + 2 {a_6 \over a_2}
{m_{D_s}^2 \over (m_b - m_c) (m_c + m_s)} \right| = 10.9 \% \\
\cr
{\rm For} \hs{3mm} \bar{B}^0 \ra D^{+}D_s^{*-} ; \hs{10mm} &&
\left|{\cA_P \over \cA_T} \right| =
\left|{a_4 \over a_2} \right| =  4.5 \%
\eeqa
Here we have used the values for $m_c = 1.45$ $GeV$, and $m_s = 170$ $MeV$.
The penguin contributions for $B \ra DD_s$ is  two times larger
than that in $B \ra DD_s^{*}$, and is a sizable effect.
This results agrees well with that of Ref.\cite{GKYP}.
Hence penguin contributions can affect the value of the decay constants
$f_{D_s}$ and $f_{D^{*}_s}$.

From (\ref{cb11a}), (\ref{cb12a}), (\ref{pb9}) and (\ref{pb10})
we obtain the following results including penguin contributions :
\begin{eqnarray}
f_{D_s^*}=336\pm 79\ {\rm MeV},\qquad f_{D_s}=219\pm 46\ {\rm MeV}& &
{\rm for\ (pole/pole)},
\nonumber\\
f_{D_s^*}=336\pm 79\ {\rm MeV},\qquad f_{D_s}=239\pm 50\ {\rm MeV}& &
{\rm for\ (pole/const.)}.
\label{c22p}
\end{eqnarray}
From (\ref{c23}), (\ref{pb9}) and (\ref{pb10})
the ratio of the vector and pseudoscalar decay constants
$f_{D_s^*}/f_{D_s}$ is given by
\begin{equation}
{f_{D_s^*}\over f_{D_s}}=1.36 \cdot
{|V_{cb}\, F_0^{BD}(m_{D_s}^2)|\over |V_{cb}\, F_1^{BD}(m_{D_s^{*}}^2)|}
\cdot
\left[
{{\cal{B}}({\bar{B}}^0\rightarrow D^+ D_s^{-*})\over
{\cal{B}}({\bar{B}}^0\rightarrow D^+ D_s^{-})} \right]^{{1/2}}
\cdot \left({0.8951\over 0.9582}\right)
\label{c23p}
\end{equation}
which gives
\begin{eqnarray}
{f_{D_s^*}\over f_{D_s}}=1.53\pm 0.45& &{\rm for\ (pole/pole)},
\nonumber\\
{f_{D_s^*}\over f_{D_s}}=1.41\pm 0.41& &{\rm for\ (pole/const.)}.
\label{c24p}
\end{eqnarray}

The decay constant depends the $q^2$ behaviour of the
form factor $F_0(q^2)$. However the amount of change is less than $10 \%$
as shown in (\ref{c22p}). This shows that the decay constant is not
strongly dependent on the behaviour of the form factor.
 As shown in Table \ref{table5},
when we include the QCD-penguin contributions,
the value of the decay constant $f_{D_s^{*}}$ is increased by $4 \%$; however,
for $f_{D_s}$ it is increased by up to $12 \%$, and the ratio 
$f_{D_s^{*}}/f_{D_s}$ is  decreased by $6 \%$.
Table 3 shows that our result for $f_{D_s}$ agrees well with others' predictions
 and experimental data,
except that  of Hwang and Kim \cite{hk1, hk2}. 
For the ratio  $f_{D_s^{*}}/f_{D_s}$,
our results have a value greater than 1. On the contrary, 
Browder et al. \cite{browder}
and Hwang and Kim 2 \cite{hk2} have a value less than 1, which is quite
strange, because their results show that the decay constant of the vector meson 
is smaller than
that of the psudoscalar meson with the same quark contents.
In fact, the decay constant of $\rho$ meson is 1.5 times greater than that of
$\pi$ meson.
By measuring the ratio $f_{D_s^{*}}/f_{D_s}$, we can check directly
their analysis and methods. 

\vs{7mm}
{\bf \large 5. Conclusion}

Within the factorization approximation, we have calculated the weak decay constants
$f_{D_s}$ and $f_{D_s^{*}}$ from the semileptonic decay,
$\bar{B}^{0} \ra D^{+}\ell^{-} \nu$ and the hadronic decay modes,
$\bar{B}^{0} \ra D^{+}\bar{D}_s^{(*)}$. We have also considered the effect of two
different $q^2$-dependences of the form factor $F_0^{BD}(q^2)$.
The value of $f_{D_s}$ is  altered by less than $10 \%$ for 
different form factors.
In our analysis, we also considered the QCD-penguin contributions in hadronic two body 
decays
within the NDR renormalization scheme in a next-to-leading order calculation.
The penguin effects for $B \ra DD_s$ decay are quite sizable. We obtain
$f_{D_s} = 219 \pm 46$ $MeV$ for the monopole type of $F_0^{BD}$,
$f_{D_s} = 239 \pm 50$ $MeV$ for the constant $F_0^{BD}$. We obtain
$f_{D_s^{*}} = 336 \pm 79$ $MeV$ for the ${{D}_{s}}^{*}$ meson. 
These predicted values will be improved
vastly when the large accumulated data samples is available at Belle and BaBar
experiment in near future.

\vspace*{1.0cm}

\noindent
{\em Acknowledgements} \\
\indent
The author wishes to thank D. S. Hwang for helpful discussions.
He is grateful to A. N. Kamal for reading this manuscript carefully.
Y.-Y. K. would like to thank M. Kobayashi 
for his hostitality and encouragement.
This work is supported
in part by the Basic Science Research Institute Program,
Ministry of Education, Project No. BSRI-97-2414,
and in part by the Grant-in Aid for Scientific
from the Ministry of Education, Science and Culture, Japan.\\

\pagebreak
\begin{table}[b]
\vspace*{0.5cm}
\hspace*{-1.0cm}
\begin{tabular}{|c|c|c|c|c|}   \hline
        &${\cal{B}} ({\bar{B}}^0\rightarrow D^+ \rho^-)$
        &${\cal{B}} ({\bar{B}}^0\rightarrow D^+ K^{-*})$
        &${\cal{B}} ({\bar{B}}^0\rightarrow D^+ \pi^-)$
        &${\cal{B}} ({\bar{B}}^0\rightarrow D^+ K^-)$
\\
        &$\times 10^3$&$\times 10^4$&$\times 10^3$&$\times 10^4$
\\   \hline
(pole/pole)    &$9.01\pm 1.54$&$4.62\pm 0.79$
               &$3.58\pm 0.61$&$2.74\pm 0.47$ \\
(pole/const.)  &$9.01\pm 1.54$&$4.62\pm 0.79$
               &$3.57\pm 0.61$&$2.71\pm 0.46$ \\ \hline
Experiments    &$8.4\pm 1.6\pm 0.7$&---
               &$3.1\pm 0.4\pm 0.2$&--- \\
\hline
\end{tabular}
\caption{The obtained values of the branching ratios and existing
experimental values.
\label{table3}}
\end{table}

\begin{table}[ht]
\vspace*{0.5cm}
\begin{center}
\begin{tabular}{|c||c|c|}   \hline
Coefficients  & Real Part & Imaginary Part \\
\hline
$C_1^{eff}$ & - 0.313 & 0.0 \\ \hline
$C_2^{eff}$ &   1.150 & 0.0 \\ \hline
$C_3^{eff}$ &   2.20 $\cdot 10^{-2}$ & 5.13 $\cdot 10^{-3}$ \\ \hline
$C_4^{eff}$ &  -5.11 $\cdot 10^{-2}$ & -1.54 $\cdot 10^{-2}$ \\ \hline
$C_5^{eff}$ &   1.50 $\cdot 10^{-2}$ & 5.13 $\cdot 10^{-3}$ \\ \hline
$C_6^{eff}$ &  -5.97 $\cdot 10^{-2}$ & -1.54 $\cdot 10^{-2}$ \\ \hline
\hline
\hline
$a_1$ &  0.071& 0.0 \\ \hline
$a_2$ &  1.046 & 0.0 \\ \hline
$a_3$ &  4.97 $\cdot 10^{-3}$ & 0.0 \\ \hline
$a_4$ &  -4.38 $\cdot 10^{-2}$ & -1.37$\cdot 10^{-2}$ \\ \hline
$a_5$ &   -4.90 $\cdot 10^{-3}$ & 0.0 \\ \hline
$a_6$ &  -5.47 $\cdot 10^{-2}$ & -1.37 $\cdot 10^{-2}$ \\ \hline
\hline
\end{tabular}
\end{center}
\caption{The values of the effective wilson coefficient $C_{i}^{eff}$
and $a_{i}$ with the $\mu = m_b = 5.0$ $GeV$, $m_c = 1.45$ $GeV$
in NDR scheme at NLO calculation. The coefficients $a_{2i}$ and $a_{2i-1}$
are defined by $a_{2i-1} = C_{2i-1}^{eff} + C_{2i}^{eff}/N_c$, 
$a_{2i} = C_{2i}^{eff} + C_{2i-1}^{eff}/N_c$ and we have taken $N_c = 3$. 
\label{table4}}
\end{table}
\begin{table}[ht]
\vspace*{0.5cm}
\begin{center}
\begin{tabular}{|c|c|c|c|}   \hline
        &$f_{D_s^*}$ (MeV)&$f_{D_s}$ (MeV)&$f_{D_s^*}/f_{D_s}$ \\
\hline \hline
(pole/pole) at Tree level   &$322\pm 76$&$196\pm 41$&$1.64\pm 0.48$ \\
(pole/const.) at Tree level &$322\pm 76$&$214\pm 45$&$1.51\pm 0.44$ \\
\hline
(pole/pole) with Penguin   &$336\pm 79$&$219\pm 46$&$1.53\pm 0.45$ \\
(pole/const.) with Penguin &$336\pm 79$&$239\pm 50$&$1.41\pm 0.41$ \\
\hline \hline
Browder $et$ $al.$ \cite{browder}    &$243\pm 70$&$277\pm 77$
               &$0.88 \pm 0.35$ \\ \hline
Hwang and Kim 1 \cite{hk1}&$362\pm 15$&$309\pm 15$&$1.17\pm 0.02$ \\
Hwang and Kim 2 \cite{hk2}&$336\pm 13$&$399\pm 24$&$0.84\pm 0.03$ \\ \hline
Capstick and Godfrey \cite{cg}&&$290\pm 20$& \\
Dominguez \cite{doming}&&$222\pm 48$& \\
UKQCD \cite{UKQCD}&&$212^{+4+46}_{-3-7}$& \\
BLS \cite{BLS}&&$230\pm 7\pm 35$&\\
MILC \cite{MILC}&&$199\pm 8^{+40+10}_{-11-0}$&\\
\hline
WA75 \cite{wa75}&
&$238 \pm 47 \pm 21 \pm 48 $& \\
CLEO 1 \cite{cleo94} &
&$282 \pm 30 \pm 43 \pm 34 $& \\
CLEO 2 \cite{cleo95} &
&$280 \pm 19 \pm 28 \pm 34 $& \\
BES \cite{bes} &
&$430^{+ 150}_{-130} \pm 40 $& \\
E653 \cite{E653} &
&$190 \pm 34 \pm 20 \pm 26 $& \\
\hline
\end{tabular}
\end{center}
\caption{The obtained values of
$f_{D_s^*}$ (MeV) and $f_{D_s}$ (MeV), and their ratio
$f_{D_s^*}/f_{D_s}$, and the results from other theoretical
calculations and existing experimental results.
Here we refered the corrected $f_{D_s}$ values \cite{cleo98}
for the experimental data \cite{wa75} - \cite{E653}.
\label{table5}}
\end{table}

\end{document}